\documentclass{ws-mpla-hep}

\begin{document}
\def\preprint{}

\markboth{Federico A. Ceccopieri}
{Trento proceedings}
\catchline{}{}{}{}{}

\title{SCALE EVOLUTION OF UNINTEGRATED DISTRIBUTIONS \\AND 
THE  $p_t$-SPECTRUM OF GAUGE BOSONS}  

\author{\footnotesize FEDERICO ALBERTO CECCOPIERI}

\address{Vrije Universiteit Brussels, \\
Interfaculty Institute for High Energy (IIHE)\\ 
Pleinlaan 2, 1040 Bruxelles,  Belgium \\
federicoalberto.ceccopieri@cern.ch}

\maketitle

\pub{Received (Day Month Year)}{Revised (Day Month Year)}

\begin{abstract}
We present predictions for the $Z$-boson $p_t$-spectrum at Tevatron
within the framework of unintegrated distributions evolved according 
to evolution equations recently proposed by us. 
We discuss the dependence of the results on the choice of non-perturbative parameters, the coupling
constant and the impact of soft gluon resummation.

\keywords{TMD distributions, Drell-Yan, pQCD, evolution equations, transverse momentum resummation}

\end{abstract}

\ccode{PACS Nos.: 12.38.Bx,12.38.Cy,13.60.-r,13.85.Ni}

\section{Introduction}	

Transverse momentum dependent, or equivalently, unintegrated distributions are currently 
object of intense research activity. The motivation for such an interest 
relies on their wide range of applicability, from spin physics to jet observables
in high energy collisions at LHC.  Their correct formalization in quantum field theory, 
in the present case in quantum chromodynamics, is however far from being trivial\cite{CS}. 
In order to investigate various properties of unintegrated distributions, 
detailed calculations in various gauges have been performed\cite{Stefanis,Hautmann}.
The structure of additional rapidity divergences is understood
and \textit{ad hoc} subtraction scheme has been proposed\cite{Hautmann}.
Their factorization properties in hard processes have been also investigated
and a factorization theorem has been given\cite{Ji}.  
A point which, to date, escapes a rigorous 
answer is how these distributions behave as long as the scale which characterizes the hard process is varied.
If this behaviour were known we would be able to relate to each other results coming from 
different experiments, possibly at different energies. The latter possibility is therefore of great 
phenomenological  importance.
Although a definitive answer to this question, especially in the light of new developments in the field, 
is absent in the literature, there have been however some attempts.  In particular     
evolution equations for unintegrated distributions  were proposed in the unpolarized time-like case\cite{BCM} and 
very recently extended to space-like kinematics\cite{CT}.
In a subsequent phemonelogical study\cite{CT2}, performed in the context of semi-inclusive deep inelastic scattering, 
it was shown that a reasonable description of data could be obtained once unintegrated evolution equations 
were solved with suitable, but motivated, initial conditions and assuming factorization for the cross-sections of interest. 
This result has stimulated us to apply the same formalism to Drell-Yan type process in hadronic collisions.
The $p_t$-spectrum of the gauge boson has, in fact, a rich 
structure and manifests many perturbative and non-perturbative features of the underlying theory.
In particular techniques for the resummation of the perturbative series in the multiple soft gluon emission limit 
were first developed for this prototype observable\cite{DDT,PP,CSS}.

\section{Unpolarized evolution}	
\begin{figure}[t]
\centerline{\psfig{file=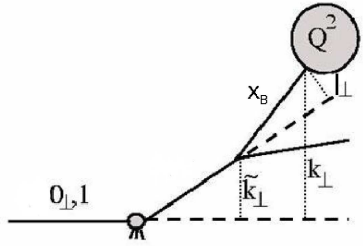,width=2.0in}}
\vspace*{8pt}
\caption{The incoming proton on the bottom left is assigned to have unitary 
momentum and defines the reference axis respect to which the transverse momentum 
$\bm{\widetilde{k}}_{\perp}$ and $\bm{k}_{\perp}$ are defined. 
The small blob represents the iteration of emissions in the parton ladder of whose 
only the last, $\widetilde{k} \rightarrow k +p $, is explicitely shown. 
$\bm{l}_{\perp}$ is the relative transverse momentum between parton $k$ and $p$. 
The interacting parton $k$ enters the hard scattering vertex indicated by the blob on top
of the diagram.\protect\label{fig1}}
\end{figure}

We briefly summarize the basic ingredients of $k_t$-evolution equations. 
Let us consider parton emissions off a active, space-like, parton line in ladder approximation. 
In the collinear limit, at each branching, the active parton increases its virtuality 
and acquires a small relative transverse momentum with respect to the parent.
These iterated emissions  generate therefore an appreciable transverse momentum, 
up to the order of the hard scale in the process, which adds to the non-perturbative one due to  
Fermi motion of the parton in the parent hadron.
Collinear emissions give however leading logarithmic corrections to cross-sections
when the transverse momenta are ordered along the ladder and can be resummed to all orders
by using DGLAP evolution equations\cite{DGLAP}. 
In the unintegrated case\cite{BCM,CT}, unlike DGLAP case, the integration on relative transverse momenta at each 
branching are left undone 
\begin{multline}
\label{dglap_TMD_space}
Q^2 \frac{\partial \mathcal{F}_{P}^{i}(x_B,Q^2,\bm{k_{\perp}})}{\partial Q^2}
=\frac{\alpha_s(Q^2)}{2\pi}\int_{x_B}^1 \frac{du}{u^3} 
P_{ji}(u,\alpha_s(Q^2)) \cdot\\ \cdot \int \frac{d^2 \bm{l_{\perp}}}{\pi}\,\delta(\,(1-u)Q^2-l^2_{\perp})
\,\mathcal{F}_{P}^{j}\Big(\frac{x_B}{u},Q^2, \frac{\bm{k}_{\perp}-\bm{l}_{\perp}}{u} \Big)\,,
\end{multline}
as indicated by the additional $d^2 \bm{l_{\perp}}$ integration. 
In the splitting $\widetilde{k} \rightarrow k +p$ depicted in Fig.~(\ref{fig1}), 
the parton $k$ carries a fraction $u$ of the fractional momentum of $\widetilde{k}$.
With this notations the following mass-invariant constraint can be derivered:
\begin{equation}
\label{mass}
l^2_{\perp}=-(1-u)k^2+u(1-u)\widetilde{k}^2-u p^2\,.
\end{equation}
If one assumes that the virtualities increase along the ladder, $k^2\gg\widetilde{k}^2$, 
and on-shell partons are emitted,  $p^2=0$, the last two terms in eq.~(\ref{mass})
can be disregarded. In these limits, setting $-k^2=Q^2$ one obtains
$l^2_{\perp}=(1-u)Q^2$, which can be found in the $\delta$-function in eq.~(\ref{dglap_TMD_space}).  
The transverse arguments of $\mathcal{F}_{P}^{i}$ on r.h.s. 
of eq.~(\ref{dglap_TMD_space}) are derivered by taking into account the Lorentz boost 
of transverse momenta from the emitting parton $\widetilde{k}$ reference frame 
to the interacting $k$ parton one\cite{GSW}. 
In particular, the transverse momentum $\bm{\widetilde{k}}_{\perp}$ of the parton which undergoes the splitting
can be expressed as follows 
\begin{equation}
\label{boost}
\bm{\widetilde{k}}_{\perp}=(\bm{k}_{\perp}-\bm{l}_{\perp})/u\,.
\end{equation}
Unintegrated parton distribution functions $\mathcal{F}_{P}^{i}(x_B,Q^2,\bm{k_{\perp}})$ 
in eq.~(\ref{dglap_TMD_space}) give
the probability to find, at a given scale $Q^2$, a parton  $i$ 
with longitudinal momentum fraction $x_B$ and transverse momentum $\bm{k}_{\perp}$ 
relative to the parent hadron, see Fig.~(\ref{fig1}).
$P_{ji}(u)$ are the space-like splitting functions. 
The unintegrated distributions fulfil the normalization condition:
\begin{equation}
\label{spacelike_norm}
\int d^2 \bm{k}_{\perp} \mathcal{F}_{P}^{i}(x_B,Q^2,\bm{k}_{\perp})=
f_{P}^{i}(x_B,Q^2)\,,
\end{equation}
where $f_{P}^{i}$ are ordinary parton distributions.
It is important to remark that, given eq.~(\ref{spacelike_norm}), 
performing the $d^2 \bm{k}_{\perp}$ integration on both side of
eq.~(\ref{dglap_TMD_space}), we recover the integrated evolution equations\cite{DGLAP} for $f_{P}^{i}$.

\section{The $p_t$-spectrum of $Z$ boson at Tevatron}
\label{sec:Zpt}
\noindent  
The analysis described in the following is inspired to a similar one\cite{Kwiecinski},
in which however a different kind of evolution for of unintegrated distributions is assumed.
The differential cross-sections for $Z$-production at rapidity $y$ and transverse momentum $\bm{p}_t$
is given by
\begin{multline}
\label{Zpt_cross}
\frac{d^4\sigma}{dy d^2 \bm{p}_t}=
\sigma_0\sum_q w_q^2 
\int d^2  \bm{k}_{\perp,1} \int d^2  \bm{k}_{\perp,2} 
\,\delta^{(2)}(\bm{k}_{\perp,1} +\bm{k}_{\perp,2}  - \bm{p}_t)\cdot \\
\cdot \Big[\mathcal{F}_q(x_1,\mu^2,\bm{k}_{\perp,1}) 
\mathcal{F}_{\bar{q}}(x_2,\mu^2,\bm{k}_{\perp,2}) + (1\leftrightarrow2)\Big]\,.
\end{multline}
This formula can be understood as the counterpart of the standard factorization formula
for Drell-Yan type processes\cite{Kwiecinski}. 
The weak charges are denoted by $w_q$, the parton momentum fractions are evaluated through 
$x_{1,2}=m_T/ \sqrt{s}\exp(\pm y)$
with $m_T=\sqrt{M^2+p_t^2}$ being the transverse mass of the gauge boson. 
The hadronic collision energy is $\sqrt{s}=1.8$ TeV and 
$\sigma_0=\frac{\pi \sqrt{2} G_F m_Z^2}{3s}$.
Parton distributions are evaluated at the scale $\mu^2=m_Z^2$. 
The differential cross-section, eq.~(\ref{Zpt_cross}), is then integrated according
to the experimental cuts\cite{D0} and accounted for the $Z$ branching ratio 
into electrons, $BR(e^{+}e^{-})=0.033632$.
\begin{figure}[t]
\centerline{\psfig{file=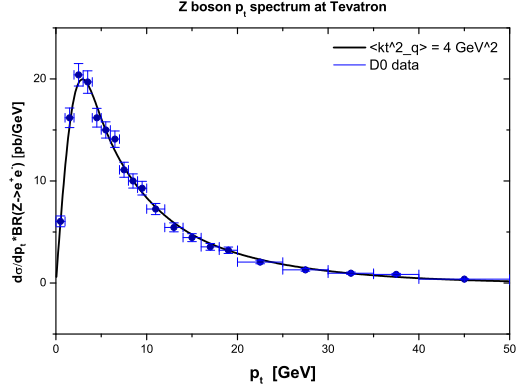,,width=3.0in}}
\vspace*{8pt}
\caption{Predictions for the $Z$-boson $p_t$-spectrum 
along with Tevatron data.\protect\label{fig2}}
\end{figure}
The initial conditions for $k_t$-evolution equation at the minimum scale $Q_0^2= 5$ Ge$V^2$ 
are chosen as a product of longitudinal parton densities\cite{GRV}
times a gaussian transverse factor 
with $x$-independent and flavour independent width\cite{CT2}:
\begin{equation}
\label{ic}
\mathcal{F}_P^i(x_B,Q_0^2,\bm{k}_{\perp})=  f_P^i(x_B,Q_0^2)
\,\frac{e^{\frac{-k_{\perp}^2}{<k_{\perp,i}^2>}}}{\pi<k_{\perp,i}^2>} \,\,\,
\;\;\; i=q,\bar{q},g\,.
\end{equation}
We evolve light flavours only\cite{GRV} and effects of heavy flavours are included in the 
running of the strong coupling evaluated in leading logarithmic approximation.
We tune the parameters appearing in the intial conditions, eq.~(\ref{ic}), as well as the strong 
coupling at the $Z$-boson mass, $\alpha_s(m_Z^2)$, to data. 
During this procedure we have observed two peculiar features: 
setting the quark intrinsic momentum $\langle k_{\perp,q}^2 \rangle$ to larger values shifts the 
position of the maximum towards higher $p_t$ while lowering the value of the coupling, 
$\alpha_s(M_Z^2)$, do overestimate its height and \textit{viceversa}.
The tuning procedure gives for the quarks (and antiquarks) 
an intrinsic momentum $\langle k_{\perp,q}^2 \rangle$=4 Ge$V^2$.
Quite interestingly the same amount of intrinsic transverse momentum is also required 
by Monte Carlo programs\cite{Seymour} when used to predict the same process.
We will discuss this large value, well above the one expected by Fermi motion,
after the discussion of soft gluon resummed results.  
Predictions are, as expected, almost insensitive to the gluon intrinsic transverse momentum 
so that we fix it at $\langle k_{\perp,g}^2 \rangle$=1 Ge$V^2$.
In order to have a reasonable agreement with data, a large value of the coupling is required,
namely $\alpha_s(M_Z^2)=0.150$.
The high value for $\alpha_s(M_Z^2)$ indicates that a more rapid evolution is necessary, 
especially in the low $p_t$ region. 
In order to have a better description of the $p_t$-distribution tail, 
we use the full invariant mass-constraint at the branching vertex, eq.~(\ref{mass}),
still however considering emission of massless partons, $p^2=0$.
The predictions within these settings are shown, along with experimental data\cite{D0}, 
in Fig.~(\ref{fig2}).
The main features already observed in the analysis\cite{CT2} of semi-inclusive deep inelastic 
scattering appear here again: the use of  $k_t$-dependent distributions 
allows one to describe gauge boson $p_t$-spectrum without 
resorting to any artificial procedure\cite{CSS} to match higher order 
pQCD corrections at large $p_t$ and non-perturbative predictions at small $p_t$, 
possibly corrected for soft gluon emissions.  This quite interesting feature 
is the result of taking in full account the transverse kinematics in the proposed evolution equations.
As is well known, the low $p_t$ part of the spectrum is sensitive not only to pure non-perturbative 
effects but also to effects coming 
from multiple soft gluon emissions.\cite{DDT,PP,CSS} In this case the $k_t$-evolution equations 
can be slightly modified in order to resum logarithms of soft nature\cite{KT}.	  
In the non-singlet channel the resummed evolution equation reads
\begin{multline}
\label{dglap_TMD_space_SGR}
Q^2 \frac{\partial \mathcal{F}_{P}^{i}(x_B,Q^2,\bm{k_{\perp}})}{\partial Q^2}
=\int_{x_B}^1 \frac{du}{u^3} \Big[ 
\frac{\alpha_s(Q^2(1-u))}{2\pi}
\widehat{P}_{qq}(u) \Big]_+ \cdot\\ \cdot \int \frac{d^2 \bm{l_{\perp}}}{\pi}\,
\delta \Big(\,(1-u)Q^2-l^2_{\perp} \Big)
\,\mathcal{F}_{P}^{j}\Big(\frac{x_B}{u},Q^2, \frac{\bm{k}_{\perp}-\bm{l}_{\perp}}{u} \Big)\,. \nonumber
\end{multline} \nonumber
In the previous equation $\widehat{P}_{qq}(u)$ denotes the unregularized splitting function.
The resummation of leading soft logarithms is performed by changing the argument 
of the running coupling from the virtuality to the relative transverse momentum 
of partons at each branching, $\alpha_s(Q^2) \rightarrow \alpha_s(l_\perp^2)$.
\begin{figure}[t]
\centerline{\psfig{file=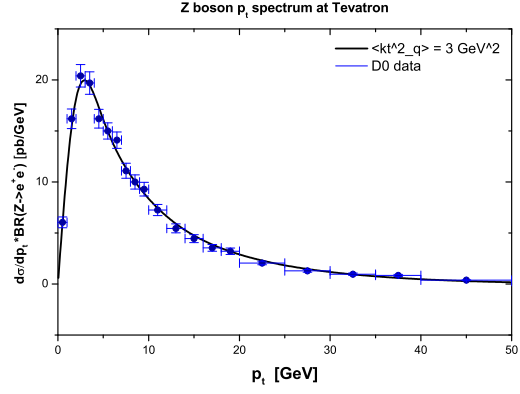,,width=3.0in}}
\vspace*{8pt}
\caption{Soft gluon resummed predictions for the $Z$-boson $p_t$-spectrum 
along with Tevatron data.\protect\label{fig3}}
\end{figure}
We wish to note that eq.~(\ref{Zpt_cross}) can be easily recast in impact parameter space\cite{Kwiecinski}.
Taking advantage of the solution of the $k_t$-evolution equations in the soft limit\cite{KT},
is then possible to recover the well known results for the perturabtive form factor in double logarithmic approximation\cite{PP}.
The off-diagonal term in the Altarelli-Parisi splitting matrix are assigned to have 
the standard coupling,  $\alpha_s(Q^2)$, since no soft enhancement is present in the $q\rightarrow g(g)$ and
$g\rightarrow q (\bar{q})$. 
The soft gluon resummation can be also performed in the gluon channel\cite{CDT} 
and the whole formalism extended up to next-to-leading logarithmic accuracy \cite{KT,CDT}. 
The latter improvements however are not still implemented. 
Despite for $u\rightarrow 1$ virtual and real contributions exactly cancel, as guaranteed by the plus prescription,
the rescaled coupling can be sensitive to how the infrared limit is approached. 
We adopt here the simplest model\cite{PP}, \textit{e.g.} a freezed coupling 
$\alpha_s(k_\perp^2 +\langle g_{\perp}^2 \rangle )$ 
being $g_{\perp}^2=0.5$ Ge$V^2$ the freezing scale. We stress that more refined 
prescription for the infrared behaviour of the strong coupling could be used\cite{APT}.  
In Fig.~(\ref{fig3}) we show the predictions which include soft gluon resummation as 
we have described above. The tuning procedure gives 
an intrinsic momentum $\langle k_{\perp,q}^2 \rangle$=3 Ge$V^2$, 
and a much lower value for the coupling, $\alpha_s(M_Z^2)=0.120$. 
Moreover the insensitivity to gluon parameters persists so that 
we still fix the gluon intrinsic momemtum to $\langle k_{\perp,g}^2 \rangle$=1 Ge$V^2$.
The inclusion of higher order terms in the perturbative calculation provided by the resummation 
has therefore the effect of strongly reduce the coupling and slightly reduce  
the quark intrinsic transverse momentum,  not still in the range of what expected by Fermi motion. 
In the resummed case, the sensitivity to the quark intrinsic momentum is shown in  Fig.~(\ref{fig4}).
\begin{figure}[t]
\centerline{\psfig{file=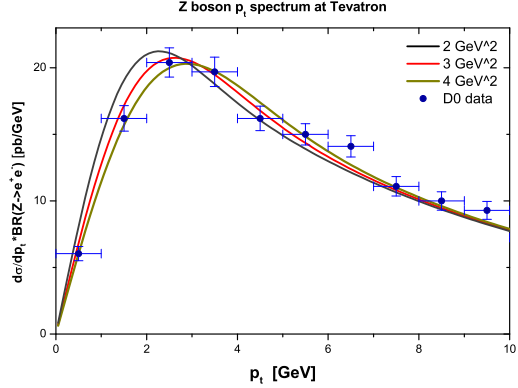,,width=3.0in}}
\vspace*{8pt}
\caption{Effects on the predicted distribution by the variation of the quark intrinsic momentum. 
Curves are shown for the following values : $\langle k_{\perp,q}^2 \rangle$=2,3,4 Ge$V^2$. 
\protect\label{fig4}}
\end{figure}
The fact that a large amount of intrinsic momentum 
is required in the description of the Z-boson data is known in the literature
and it persists even when soft resummation is pushed to next-to-leading logarithmic accuracy\cite{florence}. 
There are however attempts to solve this problem, either using $x$-dependent (\textsl{i.e.} energy dependent) 
non-perturbative form factor\cite{BLNY} or introducing modifications to the coupling constant\cite{Seymour}. 
The former approach is particularly suitable when combined analysis of low energy Drell-Yan and 
$Z$ or $W$ data are performed. A possible energy dependence of the non-perturbative form factor used 
in that works, and here embodied in the transverse part of the initial conditions, eq.~(\ref{ic}), 
is therefore of great importance in view of LHC Drell-Yan physics program.  
In order to asses the reliability of all these results a careful analysis 
of the uncertainty due to scales and parameters variation is presently under way.
Furthermore an analysis of low energy Drell-Yan is also planned 
as well as the implementation of soft gluon resummation in the gluon channel.

\section*{Conclusions}
We have found that the predictions based on the proposed evolution equations for 
the unintegrated distributions are able to reproduce Tevatron data 
on the $p_t$-spectrum of the $Z$-boson. However the unresummed 
results are characterized by large values of the quark intrinsic transverse momentum and 
coupling constant. The inclusion of soft gluon resummation strongly reduces  
the value of the coupling and brings it more close to world average. On the other hand 
it only slightly reduces the large values of the quark intrinsic
transverse momentum with respect to unresummed predictions, 
leaving therefore open the problem of the nature of the latter. 
Owing to these results and within the evolution equations scheme adopted in this analysis, 
we conclude that non-perturbative parameters differ significantly from the one used 
in our semi-inclusive DIS data analysis. 

\section*{Acknowledgments}
The author would like to thank the Organizers of the Workshop "Recent
Advances in Perturbative QCD and Hadronic Physics", ECT*, Trento (Italy),
for the invitation and would like to express his best wishes to Professor A. V. Efremov.
The author also would like to thank Luca Trentadue, Nicos Stefanis, Francesco Hautmann, Barbara Pasquini, 
Ugo Aglietti and Massimiliano Grazzini for valuable correspondence or discussions. 
A special thank goes to Oleg Teryaev for his warm interest on the subject.


\begin{thebibliography}{0}
\bibitem{CS}J.~C.~Collins, D.~E.~Soper, \textsl{Nucl.~Phys.~} \textbf{B193}  (1981) 381, 
Erratum-ibid. \textbf{B213} (1983) 545. 
\bibitem{Stefanis} I.~O.~Cherednikov, N.~G.~Stefanis, \textsl{Phys.~Rev.~} \textbf{D80} (2009) 054008; \\
\textsl{Nucl.~Phys.~} \textbf{B802}  (2008) 146; \textsl{Phys.~Rev.~} \textbf{D77} (2008)  094001. 
\bibitem{Hautmann}  J.~C.~Collins, F.~Hautmann, \textsl{JHEP} \textbf{0103}  (2001) 016; \\
F.~Hautmann, \textsl{Phys.~Lett.~} \textbf{B655} (2007) 26.
\bibitem{Ji} X.~Ji, J.~Ma, F.~Yuan, \textsl{Phys.~Rev.~}\textbf{D71} (2005) 034005.
\bibitem{BCM} A.~Bassetto, M.~Ciafaloni, G.~Marchesini, \textsl{Nucl.Phys.} \textbf{B163} (1980) 477.
\bibitem{CT} F.~A.~Ceccopieri, L.~Trentadue, \textsl{Phys.~Lett.~} \textbf{B636} (2006)  310.
\bibitem{CT2}F.~A.~Ceccopieri., L.~Trentadue, \textsl{Phys.~Lett.~} \textbf{B660} (2008) 43.
\bibitem{DDT} Y.~L.~Dokshitzer, D.~Diakonov, S.~I.~Troian, \textsl{Phys.~Rept.~} \textbf{58} (1980) 269. 
\bibitem{PP} G.~Parisi, R.~Petronzio, \textsl{Nucl.~Phys.~} \textbf{B154} (1979) 427. 
\bibitem{CSS} J.~C.~Collins, D.~E.~Soper, G.~ Sterman, \textsl{Nucl.~Phys.~} \textbf{B250} (1985) 199.
\bibitem{DGLAP}V.~N.~Gribov and L.~N.~Lipatov, \textsl{Sov. J. Nucl. Phys.} \textbf{15}  (1972) 438;\\
L.~N.~Lipatov, \textsl{Sov. J. Nucl. Phys.} \textbf{20}  (1975) 94;\\
G.~Altarelli and G.~Parisi, \textsl{Nucl. Phys.} \textbf{B126} (1977) 298;\\
Y.~L.~Dokshitzer \textsl{Sov. Phys. JETP} \textbf{46} (1977) 641.
\bibitem{GSW} S. Gieseke, P. Stephens and B. Webber, \textsl{JHEP} \textbf{12}  (2003) 045.
\bibitem{Kwiecinski}J.~Kwiecinski, A.~Szczurek, \textsl{Nucl.~Phys.~} \textbf{B680} (2004) 164.
\bibitem{GRV} M.~Gluck, E.~Reya, and A.~Vogt, \textsl{Z.~Phys.~} \textbf{C67}  (1995) 433. 
\bibitem{Seymour}S.~Gieseke, M.~H.~Seymour, A.~Siodmok, \textsl{JHEP} \textbf{06} (2008) 001.
\bibitem{D0} D0 Collaboration (B. Abbott \& al.~) \textsl{Phys.~Rev.~Lett.~} \textbf{84} (2000) 2792. 
\bibitem{KT} J.~Kodaira, L.~Trentadue, \textsl{Phys.~Lett.~} \textbf{B112} (1982) 66.
\bibitem{CDT} S.~Catani, E.~D'Emilio, L.~Trentadue, \textsl{Phys.~Lett.~} \textbf{B211} (1988)  335.
\bibitem{APT} D.~V.~Shirkov, I.~L.~Solovtsov, \textsl{Phys.~Rev.~Lett.~} \textbf{79} (1997) 1209. 
\bibitem{florence} G.~Bozzi, S.~Catani, G.~Ferrera, D.~de Florian, M.~Grazzini,
\textsl{Nucl.~Phys.~} \textbf{B815} (2009) 174.  
\bibitem{BLNY} F.~Landry, R.~Brock, P.~M.~Nadolsky, C.~P.~Yuan, 
\textsl{Phys.~Rev.~} \textbf{D67} (2003) 073016. 
\end{thebibliography}
\end{document}